\documentstyle[graphicx,float]{article}

\begin{document}
\title{A scheme for symmetrization verification}
\date{}
\author{Pedro Sancho \\ Centro de L\'aseres Pulsados, CLPU. \\ E-37008, Salamanca, Spain}
\maketitle
\begin{abstract}
We propose a scheme for symmetrization verification in two-particle
systems, based on one-particle detection and state determination. In
contrast to previous proposals, it does not follow a
Hong-Ou-Mandel-type approach. Moreover, the technique can be used to
generate superposition states of single particles.
\end{abstract}


\section{Introduction}

Non-factorizable states play a fundamental role in quantum theory.
There are two classes of non-separable states, entangled and
(anti)symmetrized ones. Recently, there has been a boom in the
theoretical and experimental study of the first ones, specially
concerning their generation, verification and potential applications
in the fields of quantum information and conceptual foundations of quantum theory. Here, we are
mainly interested into the question of verification. References
\cite{van,ple} are two good introductions to the problem.

In contrast, there are not so many studies concerning the same issue
for (anti)symmetrized states of massive particles. In quantum
theory, two identical particles whose wave functions overlap at a
given instant are indistinguishable at any subsequent time. If one
localizes one of the particles at any instant after the overlapping,
it is impossible to identify it with any of the initial ones. In
order to correctly describe this impossibility the wave function of
the two-particle system cannot be a product state (even for
non-interacting particles) but must be symmetrized in the case of
bosons and antisymmetrized for fermions.

In quantum optics, the standard test for determining identicalness, i.
e., for testing if two photons are really in indistinguishable states (and
consequently it is necessary to symmetrize the state) is based on
the Hong-Ou-Mandel (HOM) arrangement \cite{HOM}. This technique is
based on the interference of the photons at a beam splitter
\cite{Lou}.

Several authors have proposed to use the same approach for massive
particles \cite{YaN,YaS,Hom,imp}. In \cite{YaN,YaS}, mesoscopic
electron beam splitters based on electron beam lithography on a GaAs
high-mobility two-dimensional electron gas system have been
experimentally used to test the antibunching of fermions, an
unequivocal demonstration of their antisymmetrization. Several other
arrangements can be used as beam splitters for beams of massive
particles \cite{Ada,Cro}, providing also an alternative basis for
feasible massive HOM-type tests. Unfortunately, their manipulation
is much more delicate than their counterparts for light beams,
making more difficult the application of the technique to massive
particles.

The importance of having methods able to verify (anti)symmetrization
can be easily understood. For instance, in \cite{PS} it has been
suggested to study massive two-particle interferometry by
diffraction gratings. One expects novel effects when identical
particles arrive in (anti)symmetrized states. In order to
experimentally test such effects we must know in advance if the
particles incident on the diffraction grating are in
(anti)symmetrized states. If not, a negative result in the
experiment could be imputed either to the absence of the effect or
to the state of the particles (being in a product state instead of a
(anti)symmetrized one). In contrast, if we know by sure that the
particles are in a (anti)symmetrized state a negative result could
only be interpreted as a demonstration of the absence of the effect.

We propose here a scheme able to verify (anti)symmetrization in
two-particle systems, only based on the detection of one of the two
particles and the determination of the state of the surviving one.
To be concrete, if the state of the two-particle system is
(anti)symmetrized the mode distribution of the surviving particle
will be a combination of the initial distributions of the two
particles. The scheme is, in contrast to previous proposals, a test
of identicalness not based on a HOM-type approach.

As an interesting byproduct, the technique also provides an efficient
method to generate one-particle superposition states. The surviving
particle is in a superposition of the initial states of the
two particles. The technique is specially well-suited to generate
superpositions of multi-mode states peaked around very different
momentum values.

\section{The arrangement}

We consider the same arrangement as that in \cite{San}, where the
process of detection of one of the particles in a system of two
particles in multimode states is analyzed, with
special emphasis on the one-particle interferences present in the
detection pattern.

Two identical particles are prepared in the state
\begin{equation}
|2_{fg}>= \int d^3 {\bf p} \int d^3 {\bf q} f({\bf p}) g({\bf q}) \hat{a}^+_{\bf p} \hat{a}^+_{\bf q}|0>
\label{eq:uno}
\end{equation}
where $f$ and $g$ are the mode distributions, which we assume to be
normalized, $\int d^3 {\bf p} |f(p)|^2=1=\int d^3 {\bf p} |g(p)|^2$.
In Eq. (\ref{eq:uno}) $\hat{a}^+_{\bf p}$ is the creation operator
of the mode ${\bf p}$ and $|0>$ is the vacuum state. The creation
and annihilation operators obey the usual relations $[\hat{a}_{\bf
p},\hat{a}^+_{\bf q} ]_{\mp}=\delta ^3({\bf p}-{\bf q})$. In the
double sign expressions the upper one refers to bosons and the lower
one to fermions. Note that the above state is not normalized. We
shall work with a non-normalized state (see in \cite{San} how this
state is normalized), postponing the normalization to the final
stage of the calculation.

In all the above expressions we have not included the spin indexes. In order
to simplify the presentation we assume the two particles to be in
the same spin state making superfluous the presence of the indexes.
In terms of the quantum description this assumption is equivalent to have
for the spin part of the state the form $|s>_1|s>_2$, that is, a
product state with $1$ and $2$ labeling the two particles.
We could also have (for both bosons and fermions) a symmetric
state for the spin variables, $(|s>_1|s'>_2+|s'>_1|s>_2)/\sqrt{2}$.
With any of the two above choices the part of the state related to
the spatial variables must be symmetric for bosons and antisymmetric
for fermions. We do our presentation in this framework, symmetrized
states for the spatial or momentum wave functions of bosons, and
antisymmetrized for fermions. The extension to the case of particles
in antisymmetrized spin states is straightforward, as it is only
necessary to consider antisymmetric and symmetric spatial states
for, respectively, bosons and fermions.

The two-particle system now interacts with a detector. We only
consider the cases in which only one detection occurs. Moreover, we
assume that in the detection process the particle is removed, for
instance, by absorption. In order to postselect these cases one must use a
detector able to distinguish events with one- or two-detection
events (see \cite{Nuc,Gis} for an optical detector with this
property).

Next, we show that the detection process is described by the action
of the Schr\"{o}dinger field operator. From a mathematical point of
view this operator $\hat{\psi }({\bf R},t)$ represents the
annihilation of a particle at point ${\bf R}$ and time $t$
\cite{Lan, Bay}. If the initial state of the system is a n-particle
state $|n>$, the action of the operator leaves a $n-1$ particle one,
$\hat{\psi}|n> \rightarrow |n-1>$. Thus, this operator represents in
a natural way processes in which particles are removed. Intuitively,
one can identify this removal process with the destructive detection
assumed above. With this identification $\hat{\psi }({\bf
R},t)|2_{fg}>$ describes the state of the resultant one-particle
system. At this point a criticism to this identification could be
made. Being the detection process a non-unitary one, perhaps we
should use a mixture instead of a pure state to describe the final
one-particle system. We argue here favoring the pure state choice.
Our argument is based on the analogy to detection through
photomultipliers in quantum optics \cite{Lou} (see also \cite{ap}
for a detailed treatment of absorption in multiparticle massive
beams and its description in the framework of the Glauber scheme).
In this type of destructive detection the photons are absorbed by
the atoms composing the photomultiplier. The description of the
interaction between detection device and radiation is done in terms
of the positive frequency part of the electromagnetic field operator
$\hat{\bf E}^+$. If we denote by $|R>$ the initial state of the
radiation field, its final state is assumed to be $\hat{\bf E}^+
|R>$, a pure state not a mixture. From that pure state one can
evaluate the probability of transition to a state $|R_F>$, as
$|<R_F|\hat{\bf E}^+ |R>|^2$. By analogy with the extensively
verified scheme of detection by photomultipliers we assume that
after the destructive detection we have the pure state $\hat{\psi
}({\bf R},t)|2_{fg}>$. As we shall see in the following sections,
this assumption gives rise to some conclusions that can be
experimentally tested, making the assumption verifiable.

In the expression of the field operator ${\bf R}$ refers to the
point of detection. We use this notation to emphasize that we are
considering the detector to be at a fixed position in each
repetition of the experiment. We want also remark that it does not
act as a physical variable (as it is the case of ${\bf r}$) but as a
parameter. This is a necessary consistency condition for the
calculation, because in the second quantization formalism there
cannot be explicit dependence on the spatial variables, unless it is
in a parametric form \cite{Lan}.

The field operator is given by $\hat{\psi}({\bf R},t) =\int d^3 {\bf
p} \psi _{\bf p}({\bf R}) \hat{a} _{\bf p} (t)$, with $\psi _{\bf
p}$ a complete set of orthonormal stationary wavefunctions. The most
common choice for this set is that of plane waves $\psi _{\bf
p}({\bf R})=(2\pi \hbar)^{-3/2} \exp (i {\bf p}\cdot {\bf
R}/\hbar)$. The time dependence is carried by the annihilation
operator. For planes waves it is given by $\hat{a} _{\bf p} (t)=
\hat{a} _{\bf p} \exp (-iE_{\bf p} t/\hbar)$ with $\hat{a} _{\bf p}$
the annihilation operator at time $t=0$ and $E_{\bf p}={\bf p}^2/2m$
the energy.

From now on, by the matter of simplicity, we shall restrict our
considerations to stationary situations. Consequently, we can drop
the time variables from all the expressions.

\section{Mode distribution of the surviving particle}

In our case, the final state after one detection is
\begin{equation}
\hat{\psi}({\bf R}) |2_{fg}>
\end{equation}
The evaluation of the state of the surviving particle is simple
\begin{eqnarray}
\hat{\psi}({\bf R}) |2_{fg}>= \int d^3 {\bf p} (\psi _f ({\bf R}) g({\bf p}) \pm \psi _g ({\bf R}) f({\bf p})) \hat{a}_{\bf p}^+|0> \nonumber \\
=  \psi _f ({\bf R}) |1_g> \pm \psi _g ({\bf R}) |1_f>
\label{eq:tt}
\end{eqnarray}
with $\psi _f ({\bf R})=\int d^3 {\bf p} f({\bf p}) \psi _{\bf
p}({\bf R}) $ and $|1_f>=\int d^3 {\bf p} f({\bf p})\hat{a}_{\bf
p}^+|0> $.

The above state has not yet been normalized. We denote by $|1_h>$
the normalized state of the surviving particle, which is given by
$|1_h>=\hat{\psi}({\bf R}) |2_{fg}>/ N$, with
\begin{equation}
N^2({\bf R})= |\psi _f({\bf R})|^2 + |\psi _g({\bf R})|^2 \pm 2Re(\psi _f^*({\bf R}) \psi _g({\bf R}) <1_g|1_f>)
\end{equation}

Finally, the mode distribution of the surviving particle is
\begin{equation}
h({\bf p})= \frac{\psi _f ({\bf R})}{N({\bf R})}  g({\bf
p}) \pm \frac{\psi _g ({\bf R})}{N({\bf R})} f({\bf p})
\label{eq:hh}
\end{equation}
The mode distribution is neither $f$ nor $g$, but a combination of
them. The coefficients in the combination are given by some
functions depending on the values of the initial wavefunctions at
the point of detection.

Note that only in the cases where $\psi _f ({\bf R})=0$ or $\psi _g
({\bf R})=0$, the surviving particle has a mode distribution similar
to one of the initial ones $|1_g>$ or $|1_f>$. This behaviour can be easily understood. Let us consider one
of these nodal points, for instance, $\psi _g ({\bf R})=0$. Then we
know by sure that the particle detected is of the type $f$ and,
consequently, that the surviving one is of type $g$ ($h({\bf
p})=g({\bf p})$).

In the points where simultaneously we have $\psi _f ({\bf R})=0$ and
$\psi _g ({\bf R})=0$ the expression for $|1_h>$ is undefined,
reflecting the property that at these points the probability of
detection of the particles is null and it does not make sense to
speak about the surviving one.

The fact that the mode distribution of the surviving particle
becomes a combination of the initial ones is a clear manifestation
of a superposition process. This follows directly from the last term
in Eq. (\ref{eq:tt}), but can easily be visualized if we consider
the first-quantization version of the problem, where the
wavefunction is given by
\begin{equation}
\psi _h ({\bf r})=\int d^3 {\bf p} h({\bf p}) \exp (i{\bf p}\cdot {\bf r}/\hbar)
=\frac{\psi _f ({\bf R})}{N({\bf R})} \psi _g ({\bf r}) \pm \frac{\psi _g ({\bf R})}{N({\bf R})} \psi _f ({\bf r})
\end{equation}
showing the superposition of $\psi _f$ and $\psi _g$. The
wavefunction of the surviving particle is a superposition of the two
initial wavefunctions. This superposition can be physically
understood in terms of the indistinguishability of the alternatives
available to the system, such as stated by the general criterion for
quantum superpositions. This criterion states that when the
evolution of a quantum system is compatible with several
indistinguishable alternatives the state representing the system is
a superposition of them. In our case, outside the
nodal points, the detector is unable to provide information about
the mode distribution of the particle measured, making the
alternatives {\it detected particle with mode distribution
$f$} and {\it detected particle with mode distribution $g$} indistinguishable.

When the particles are not in (anti)symmetrized states the above
results do not hold. In the second-quantization formalism used in
this paper the absence of (anti)symmetrization translates into the
use of (anti)commuting creation and annihilation operators $\hat{a}$
and $\hat{b}$. The initial state (now normalized) is
\begin{equation}
|2_{fg}>_{NS}= \int d^3 {\bf p} \int d^3 {\bf q} f({\bf p}) g({\bf q}) \hat{a}^+_{\bf p} \hat{b}^+_{\bf q}|0>
\end{equation}
If we place a detector of particles of type $a$ (the reasoning for
type $b$ is similar) the state after one-detection event is
\begin{equation}
\hat{\psi }^{(a)}({\bf R})|2_{fg}>_{NS}= \psi _f ^{(a)} ({\bf R}) |1_g>^{(b)}
\end{equation}
The above expression shows that now the mode distribution of the surviving particle is identical to the initial one.

\section{The scheme}

The above results provide the basis for a scheme of
(anti)symmetrization verification between the two initial particles.
The scheme has two steps. In the first stage, the two particles
interact with the detector. We postselect all the cases of
one-detection events. Detectors able to distinguish one- and
two-detection events would greatly simplify the task. As signaled
before, in the optical case such types of detectors are available
\cite{Nuc,Gis}. In the second step, we deal with the postselected
surviving particles. We must test if the particles are in one of the
two initial states or in a superposition of them. To this end we can
use standard tomography techniques or study the spatial detection
patterns of the surviving particles. If the detectors used in the
first step are unable to discriminate between one- and two-detection
events the postselection must also be carried out in the second
stage. Then we choose the cases in which, in each repetition of the
process, there is an event in the detector and another in the device
testing the superposition. As the efficiency of the detectors of
massive particles is in general high the scheme seems to be, in
principle, an effective one. These are the general lines of the
scheme. Now we discuss in detail a particular realization that is,
in principle, viable with present technology. We present it in Fig.
1.

We first consider the stage of preparation of the state. We
consider an arrangement where each particle is emitted by a different
source. Two main types of sources have been used in the study of
(anti)bunching effects in massive particles, magnetic traps
\cite{ott,Jel} and optical lattices \cite{fol,rom}. Sometimes the
traps and lattices use previously generated Bose-Einstein
condensates (BEC), and in other occasions gases produced by
evaporative and sympathetic cooling. In \cite{Jel,fol,rom} the atoms
are released at the switch-off of the potentials retaining the atoms
in the trap or lattice. A number of atoms of the order of $10^4$ to
$10^5$ are released, making the technique useless for our purpose of
preparing two-particle systems. From our point of view, more
interesting is the method of \cite{ott}, where a weak continuous
microwave field is applied to the BEC in the magnetic trap for
output coupling an atom laser. Some atoms inside the trap are
spin-flipped and do not longer experience the magnetic potential,
leaving the trap. The number of released atoms is much smaller than
in the previous techniques. As a matter of fact, in \cite{ott} a
mean atom number very close to $2$ was reported, making possible the
individual detection of the atoms. With this method, adequate values
of the microwave fields, and synchronization of the two fields acting
on different traps it seems possible to create two-particle states.

For different parameters of the two traps (different cooling,...) we
can have particles with different momentum distributions. Another
way to generate different momentum distributions is to collimate the
released beam via delimiters of different apertures. The uncertainty
relation tell us that if the width of the aperture is $l$ the
momentum spread will be $\delta p \sim \hbar /l$.

When the mean separation between the emission of successive pairs of
particles is large enough the postselection process of pairs can be
done in a simple way by comparing the detection times in the
detector and the tomographic device.

\begin{figure}[H]
\center
\includegraphics[width=10cm,height=10cm]{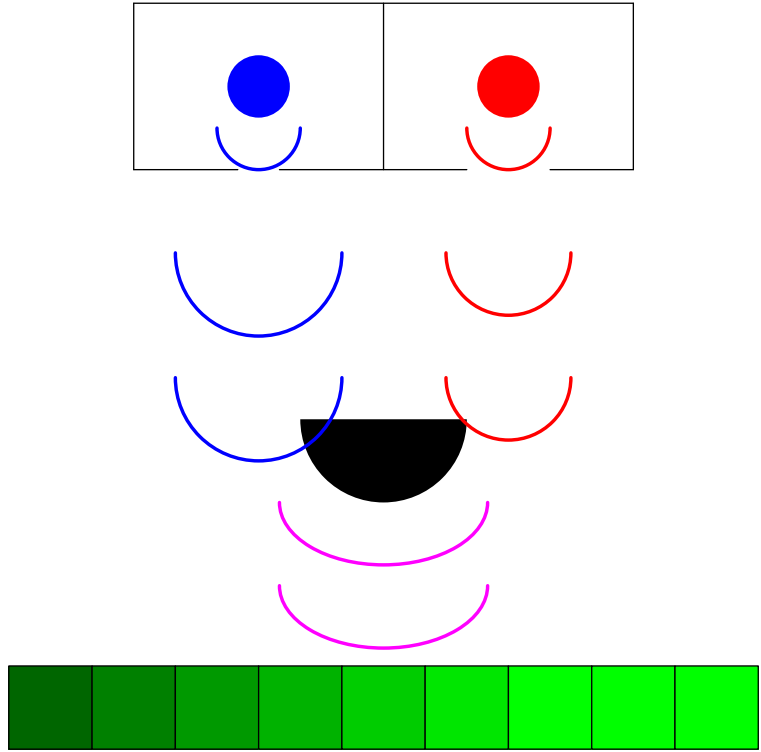}
\caption{Schematic representation of the proposed arrangement. The
boxes containing the red and blue circles represent the two sources
of particles, with typical mode distributions $f$ and $g$. The black
half-circle corresponds to the main detector. Finally, the green
boxes in the bottom of the figure represent the detectors array.}
\end{figure}

Next, we must perform tomography of the surviving particle in order
to determine its state. As it is well-known \cite{lvo}, the first
question in quantum tomography is to determine if the system is in a
pure state or it must be described by a mixture. In the first case,
measuring the detection probability in different planes along the
beam propagation direction we can determine the complete (modulus
and phase) wave function \cite{lvo}. In the second case, only the
density matrix of the system can be obtained from the measurements.
Then our first task is to determine if we are dealing with a pure
state or a mixture. In Sect. 2 we have assumed that after the
detection the two-particle system collapses into a pure one-particle
state. A possible way to assess if the surviving particle is in a
pure superposition state consists in the comparison of the spatial
detection patterns of the surviving and separately emitted
particles. To this end we place a detectors array downstream of the
main detector (that absorbing one of the particles).  With this
array we can determine the spatial detection pattern. For the
surviving particle in a pure state we have the distribution
\begin{eqnarray}
|\psi _h({\bf r}_{det})|^2 =|\alpha _f ({\bf R})|^2 |\psi _g({\bf r}_{det})|^2+ |\alpha _g ({\bf R})|^2 |\psi _f({\bf r}_{det})|^2 \nonumber \\
\pm 2 Re (\alpha _f ^*({\bf R}) \psi _g^*({\bf r}_{det}) \alpha _g ({\bf R})\psi _f({\bf r}_{det}))
\label{eq:spi}
\end{eqnarray}
with an obvious notation for the $\alpha $ coefficients, and ${\bf
r}_{det}$ denoting the location of the detectors composing the
array.

In a different run of experiments we obtain $|\psi _f ({\bf
r}_{det})|^2$, $|\psi _g ({\bf r}_{det})|^2$, $|\psi _f ({\bf R})|^2$
and $|\psi _g ({\bf R})|^2$ by considering situations where only one
of the sources emits particles. Then we compare $|\psi _h({\bf
r}_{det})|^2$ with $|\psi _f ({\bf R})|^2 |\psi _g({\bf
r}_{det})|^2+ |\psi _g ({\bf R})|^2 |\psi _f({\bf r}_{det})|^2$ in order to
see if both distributions coincide (there is no superposition of
the initial states in the surviving particle, corresponding the situation to a
mixture) or not (there is a superposition associated with a pure
state). Note that we do not need to determine the normalization
factor $N({\bf R})$ because it is not necessary to see if both distributions have the same analytical form.

With the above procedure we can determine the presence of the
superposition in the surviving particle and, consequently, if it is
in a pure state. If it is so, we can complete the tomographic
process by position measurements of the particle in different planes
after the main detector (a detectors array at each one of these planes).

It must be noted that if we only want to know if the two-particle
state was (anti)symmetrized previous to the interaction with the
main detector, it is enough to carry out the above procedure of
comparing the $|\psi _h|^2$ and $|\psi _f|^2$ and $|\psi _g|^2$
distributions, without need of a much more demanding complete
process of tomography. In effect, when the surviving particle is in
a superposition state the initial two-particle wave function
necessarily had to be in a (anti)symmetric state. The situation is
different if, in addition to determine the presence of exchange
processes in the two-particle system previous to the interaction
with the main detector, we want to know if the spatial part of the
state of the system was in a symmetric or antisymmetric state. As
discussed before, the bosonic and fermionic particles can be in a
symmetric or antisymmetric spatial state depending on the
antisymmetric or symmetric character of the spin part of the state.
The symmetric or antisymmetric character of the spatial part of the
two-particle wave function can be determined from the positive or
negative sign in the superposition of the surviving particle.
However, the value of this sign cannot be determined with the method
of comparison of $|\psi _h|^2$ and $|\psi _f|^2$ and $|\psi _g|^2$
presented in the previous paragraphs. We would need to know the
complete expression of $\psi _h$ (or equivalently, the knowledge of
$\psi _f$ and $\psi _g$ and the use of Eq. (\ref{eq:spi})). We would
need a complete tomography process.

\section{Conclusions}

We have presented in this paper a scheme for (anti)symmetrization
verification. Our proposal is not based on HOM-type arrangements.
The ideas presented here seem, in principle, accessible to
experimental scrutiny.

The scheme could also be used to generate superposition states in
one-particle systems, starting from two-particle ones. If the
initial two-particle state is (anti)symmetrized, the detection of
one of the members of the pair outside the nodal points leaves the
other in a superposition of the two initial states. We have
generated superposition via (anti)symmetrization and detection. An
interesting application of the method would be to generate a
multi-mode superposition state of two very different central
momenta. We only need (i) to prepare the initial particles with
momentum distributions $f(p)$ and $g(p)$ peaked around two values
$p_f$ and $p_g$ with $|p_f - p_g| \gg \sigma$, where $\sigma $ is
the typical width of the distributions, (ii) to get a non-negligible
overlapping of the two particles in order to have a
(anti)symmetrized state, and (iii) to postselect the cases with
one-detection events.

Finally, we must consider the possibility of applying the
scheme in quantum optics. The HOM technique with beam splitters is
very efficient for optical systems, making superfluous the existence
of other verification schemes. Nevertheless, our proposal could be
interesting to generate superpositions of one-photon states. A
simple calculation shows that a result similar to that obtained for the
distribution of the surviving massive particle holds in quantum
optics. Moreover, in this field there is much experience in dealing
with two-photon states, even in multimode states \cite{Lou}.


\begin{thebibliography}{99}
\bibitem{van} van Enk, S. J., L\"{u}tkenhaus, N. and Kimble H. J., Phys. Rev. A  {\bf 75}, 052318 (2007).
\bibitem{ple} Plenio, M. B. and Virmani, S., Quant. Infor. Comput. {\bf 7}, 1 (2007).
\bibitem{HOM} Hong, C. K., Ou, Z. and Mandel, L., Phys. Rev. Lett. {\bf 59} 2044 (1987).
\bibitem{Lou} Loudon, R., "The Quantum Theory of Light" (Oxford Science Publications, Oxford, 2000).
\bibitem{YaN} Liu, R. C., Odom, B., Yamamoto, Y. and Tarucha, S., Nature {\bf 391}, 263 (1998).
\bibitem{YaS} Oliver, W. D., Kim, J., Liu, R. C. and Yamamoto, Y., Science {\bf 284}, 299 (1999).
\bibitem{Hom} Bose, S. and Home, D., Phys. Rev. Lett. {\bf 88}, 050401 (2002).
\bibitem{imp} Lim, Y. L. and Beige, A., New J. Phys. {\bf 7}, 155 (2005).
\bibitem{Ada} Adams, C. S., Sigel, M. and Mlynek, J., Phys. Rep. {\bf 240} 143 (1994).
\bibitem{Cro} Croning, A. D., Schmiedmayer, J. and Pritchard, D. E., Rev. Mod. Phys. {\bf 81}, 1051 (2009).
\bibitem{PS} Sancho, P., J. Phys. B {\bf 43}, 065504 (2010); ibid, Phys. Rev. A {\bf 82} 033814 (2010).
\bibitem{San} Sancho, P., J. Phys. A {\bf 37}, 11003 (2004).
\bibitem{Nuc} Golovin, V. and Saveliev, V., Nucl. Instr. Meth. Phys. Res. {\bf 518}, 560 (2004).
\bibitem{Gis} Eraerds, P., Legr\'e, M., Rochas, A., Zbinden, H. and Gisin, N., Opt. Exp. {\bf 15}, 14539 (2007).
\bibitem{Lan} Landau, L. D. and Lifshitz, E. M., "Quantum Mechanics" (Addison-Wesley, 1965).
\bibitem{Bay} Baym, G., "Lectures on Quantum Mechanics" (W. A. Benjamin, 1969).
\bibitem{ap} Sancho, P., Ann. Phys. {\bf 323}, 1271 (2008).
\bibitem{ott} \"{O}ttl, A., Ritter, S., K\"{o}hl, M. and Esslinger, T., Phys. Rev. Lett. {\bf 95}, 090404 (2005).
\bibitem{Jel} Jeltes, T., McNamara, J. M., et al, Nature {\bf 445}, 402 (2007).
\bibitem{fol} F\"{o}lling, S., Gerbier, F., Widera, A., Mandel, O., Gericke, T. and Bloch, I., Nature {\bf 434}, 481 (2005).
\bibitem{rom} Rom, T., Best, Th., van Oosten, D., Schneider, U., F\"{o}lling, S., Paredes, B. and Bloch, I., Nature {\bf 444}, 733 (2006).
\bibitem{lvo} Lvovsky, A. I. and Raymer, M. G., Rev. Mod. Phys. {\bf 81}, 299 (2009).
\end{thebibliography}
\end{document}